\documentclass[aps,prl,reprint,superscriptaddress]{revtex4-2}

\usepackage{graphicx}
\usepackage{dcolumn}
\usepackage{bm}
\usepackage{lineno}
\usepackage{ulem}
\usepackage{soul}
\usepackage{color}
\usepackage{array}
\usepackage{booktabs}
\usepackage{multirow, makecell}
\usepackage{physics}
\usepackage{lineno}
\usepackage[colorlinks=true, allcolors=blue]{hyperref}%
\usepackage{xcolor}
\usepackage{gensymb}

\begin{document}

\title{Ultralow-Barrier Ion Transport in the Subnitride Electride $\mathrm{Ba_3N}$ for High-Rate Sodium Storage}

\author{Seulbi Kim}
\affiliation{Department of Materials Science and Engineering, Kyung Hee University, Yongin 17104, Republic of Korea}

\author{Bo Gyu Jang}
\email{bgjang@khu.ac.kr}
\affiliation{Department of Materials Science and Engineering, Kyung Hee University, Yongin 17104, Republic of Korea}

\date{\today}

\begin{abstract}
Electrides, ionic crystals in which excess electrons are delocalized within interstitial voids rather than bound to specific atomic sites, exhibit exotic physicochemical behaviors, yet their potential as electrochemical energy-storage hosts remains largely unexplored. Here, using comprehensive first-principles calculations, we demonstrate that the quasi-one-dimensional subnitride electride Ba$_3$N offers a compelling route to overcome the long-standing performance trade-offs in sodium-ion battery anodes. Spontaneous Na intercalation into the open interchain channels establishes thermodynamically stable phases, generating an exceptionally flat low-potential plateau that avoids the high-potential sloping losses of hard carbon while preserving an essential safety margin against dendrite formation. Crucially, the itinerant interstitial anionic electron sea dynamically flattens the potential landscape by suppressing site-specific orbital interactions, enabling ultrafast Na ion transport with an exceptionally low migration barrier. This work establishes a design paradigm for harnessing electride chemistry to achieve ultrafast battery electrodes.
\end{abstract}

\maketitle
\begin{figure}
\includegraphics[width=1\linewidth]{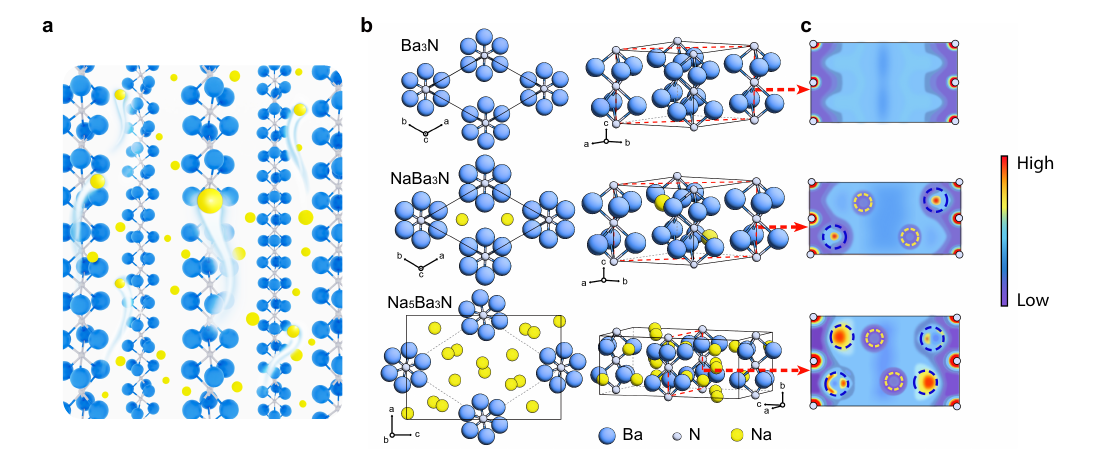}
\caption{Crystal structures and charge density distributions of Na$_x$Ba$_3$N series. (a) Schematic illustration showing Na atoms intercalated between the one-dimensional Ba$_3$N chains that are shared across all three compositions. (b) Crystal structures of Ba$_3$N, NaBa$_3$N, and Na$_5$Ba$_3$N (top to bottom), each shown as a top view (left) and a side view (right). Ba, N, and Na atoms are represented by blue, gray, and yellow spheres, respectively. (c) Calculated charge density distributions on the (110) plane for Ba$_3$N and NaBa$_3$N, and on the (002) plane for Na$_5$Ba$_3$N, with the color scale ranging from low (purple) to high (red) density. } 
\label{struct}
\end{figure}

\section{Introduction}
Sodium-ion batteries (SIBs) have gained substantial traction as a cost-effective and sustainable alternative to lithium-ion batteries (LIBs), particularly for large-scale energy storage systems where economic viability and resource abundance are paramount~\cite{yabuuchi2014}. Although SIBs share similar rocking-chair chemistry and manufacturing compatibility with LIBs, the practical implementation of SIBs has been largely hindered by the development of suitable anode materials. While diverse cathode materials—such as transition metal layered oxides, polyanionic frameworks, and Prussian blue analogues—have demonstrated promising electrochemical performance~\cite{komaba2009, komaba2010, liu2023, hurlbutt2018}, the identification of robust negative electrodes with high reversible capacity and low operating voltage remains a critical bottleneck.

In commercial LIBs, graphite is the standard anode material owing to its stable staging intercalation, high capacity (372 mAh g$^{-1}$), and robust cyclability. However, in SIBs, graphite exhibits a negligible sodium storage capacity due to unfavorable intercalation thermodynamics and the larger ionic radius of Na$^+$ (1.02 \AA\ vs. 0.76 \AA\ for Li$^+$)~\cite{moriwake2017}. To circumvent this, non-graphitizable hard carbon (HC) has been widely adopted as the benchmark anode for SIBs~\cite{stevens2000high, wang2024recent}. Nonetheless, HC typically suffers from an undesirable sloping voltage profile at high potentials (leading to degraded energy density), severe sluggish ion-diffusion kinetics and safety concerns during low-potential plateaus (risk of Na dendrite plating), and inherently low initial Coulombic efficiency (ICE) caused by irreversible Na$^+$ trapping at extensive structural defects~\cite{komaba2011, bommier2015new, qi2019slope, li2017mechanism, xiao2018low, zhang2020extended}. Alternative anode materials—including alloying-type (e.g., Sn, P, Sb) and conversion-type compounds—deliver ultrahigh theoretical capacities but are severely crippled by massive volume expansion, mechanical pulverization, and rapid capacity fading upon cycling~\cite{jung2014,li2015,jung2016,ni2018, yabuuchi2014}.

Consequently, developing fundamentally new classes of anode materials that can overcome the intrinsic trade-offs among capacity, structural integrity, and charge-transport kinetics is essential. 
Electrides are a unique class of ionic crystalline compounds in which excess electrons occupy interstitial sites, weakly confined by the electrostatic potential of the surrounding ionic framework, thereby forming an anionic electron cloud that compensates the positive charge of the cationic lattice~\cite{dawes1986first, dye1990, li2003inorganic, redko2005design, dye2009, lee2013, liu2020electrides, hosono2021advances, zhou2024van}. 
These loosely bound anionic electrons present unprecedented physicochemical properties, including ultra-low work functions and exceptionally high electrical conductivity~\cite{liu2020electrides, hosono2021advances, zhou2024van}.
Moreover, the interstitial voids occupied by these anionic electrons are expected to serve as favorable pathways for ion transport, analogous to how two-dimensional interlayer spaces in layered materials enable reversible ion intercalation and storage.
Depending on how these anionic electrons connect throughout the lattice, electrides can be classified into zero-, one-, two-, and three-dimensional systems, each forming a distinct network of interstitial channels~\cite{zhu2019}.

Among various candidates, the subnitride electride Ba$_3$N stands out as a particularly compelling host. Experimentally, the ternary subnitride series Na$_x$Ba$_3$N has been successfully synthesized across multiple stoichiometries, namely Ba$_3$N, NaBa$_3$N, and Na$_5$Ba$_3$N~\cite{rauch1992, snyder1995, stein1998}. 
Across all three compositions, Ba and N atoms consistently form an infinite one-dimensional [NBa$_{6/2}$] chain through strong ionic Ba-N bonding, into which Na atoms are progressively intercalated with increasing $x$~\cite{rauch1992, snyder1995, stein1998, zhang2022}.
Building on this experimentally verified structural family, Ba$_3$N and its Na-intercalated derivatives have further been established as three-dimensional electrides, in which anionic electrons donated by the positively charged [Ba$_3$N]$^{3+}$ chains are confined within the three-dimensional interstitial space rather than a lower-dimensional subspace~\cite{zhang2022, weaver2025}.

In this work, we systematically investigate the electrochemical sodium-storage properties and underlying ion-transport mechanisms of the one-dimensional subnitride electride Ba$_3$N using first-principles calculations. Our calculations demonstrate that spontaneous Na intercalation yields thermodynamically stable intermediate phases, producing a flat, low-potential operating plateau. Crucially, the host framework maintains robust metallic conductivity and facilitates ultrafast cation diffusion through its open interchain channels, underpinned by an exceptionally low migration barrier. Charge density analyses reveal that the spatially delocalized interstitial electron clouds play a decisive role in flattening the diffusion landscape by suppressing rigid, site-specific orbital interactions, enabling ultrafast ion transport. Notably, these compelling electrochemical features are not merely limited to Ba$_3$N, but reflect the intrinsic advantages of interstitial anionic electrons, highlighting inorganic electrides as an intriguing and largely unexplored material platform for advanced energy storage.

\section{Result}

\begin{figure*}
	\includegraphics[width=1.0\textwidth]{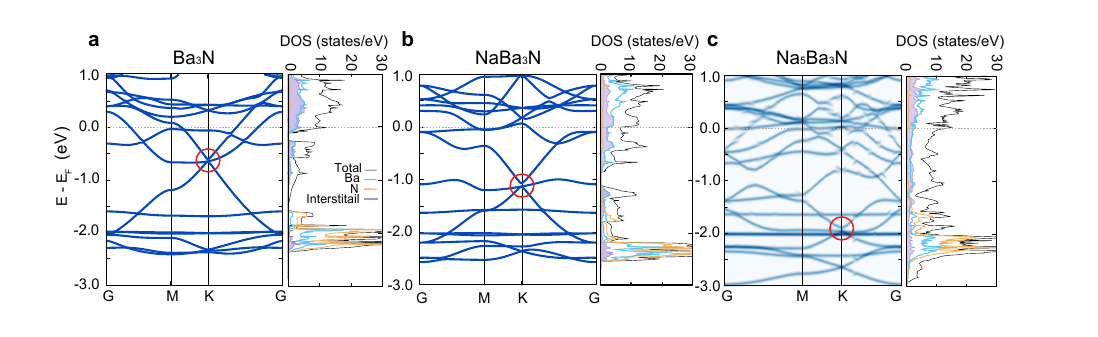}
	\caption{Band structure and density of states (DOS) of (a) Ba$_3$N, (b) NaBa$_3$N, and (c) Na$_5$Ba$_3$N. The band structure of orthorhombic Na$_5$Ba$_3$N is unfolded back to the hexagonal Brillouin zone for a direct comparison. Red circles mark the characteristic band crossing at the K point, and the DOS panels decompose the total density of states (black) into Ba (orange), N (cyan), and interstitial (purple) contributions. }
	\label{dos}
\end{figure*}

Figure~\ref{struct} illustrates the crystal structures of the Na$_x$Ba$_3$N ($x=0,1,5$) series. The foundational structural unit shared across all three phases is the infinite one-dimensional (1D) Ba$_3$N chain, which consists of face-sharing NBa$_6$ octahedra, with Na atoms progressively occupying the interchain interstitial spaces as $x$ increases (Figure~\ref{struct}a). 
While Ba$_3$N and NaBa$_3$N crystallize in the hexagonal phase ($P6_3/mcm$), Na$_5$Ba$_3$N adopts the orthorhombic phase ($Pnma$) owing to a slight distortion of Ba$_3$N chain and the distribution of intercalated Na atoms~\cite{rauch1992, snyder1995, stein1998, zhang2022}. Nevertheless, the Ba$_3$N strands are fundamentally arranged in a hexagonal lattice array across all compositions, forming an open framework with spacious interchain interstitial channels as shown in Figure~\ref{struct}b.

Due to the large electronegativity difference between Ba and N, the chemical bonding within these 1D strands approaches the ionic limit. The experimental intra-chain Ba-N bond length of 2.73 \AA\ is in close proximity to the sum of the ionic radii of Ba$^{2+}$ (1.35 \AA) and N$^{3-}$ (1.46 \AA), reflecting strong heteropolar ionic bonding inside the chains~\cite{rauch1992, snyder1995, stein1998}. Consequently, the 1D framework behaves as a positively charged ionic backbone, formally represented as [(Ba$^{2+}$)$_3$N$^{3-}$]$^{3+}$. The excess valence electrons from Ba atoms are released from the chain and confined within the interstitial spaces between the chains, leading to a [Ba$_3$N]$^{3+} \cdot 3e^-$ configuration~\cite{rauch1992, snyder1995, stein1998, oliv2005, zhang2022}. These delocalized interstitial electrons give rise to the intrinsic electride nature of the Ba$_3$N framework, which is well captured by the charge density distribution. 

Figure~\ref{struct}c shows the calculated charge density distribution of Na$_x$Ba$_3$N near the Fermi level ($E_F$). While the maximum density (red) appears near the atomic sites, substantial electron density is delocalized throughout the interstitial voids, with noticeably higher intensity at the void centers (sky blue) than near the chain boundaries (purple). Upon intercalation, sodium atoms systematically occupy the interchain interstitial voids while preserving both the rigid $\text{Ba}_3\text{N}$ framework and the delocalized 3D electride channel network~\cite{zhang2022, weaver2025}.

\begin{figure}
	\includegraphics[width=1\linewidth]{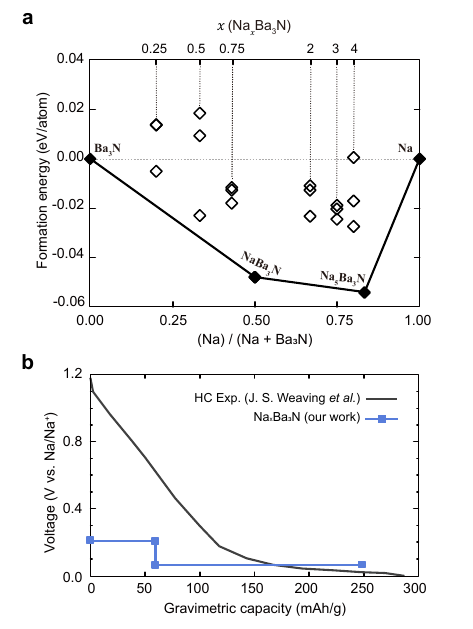}
	\caption{(a) Convex hull of the Na$_x$Ba$_3$N series. (b) Comparison of the voltage–capacity profiles of hard carbon and Ba$_3$N as a function of gravimetric capacity. The experimental data for hard carbon are taken from Ref.~\cite{weaving2020}.
	}
	\label{convex}
\end{figure}

Figure~\ref{dos} presents the calculated band structures and corresponding density of states (DOS) of the Na$_x$Ba$_3$N series. 
Starting with pristine Ba$_3$N (Figure~\ref{dos}a), the partial DOS (PDOS) of the N atom is fully occupied well below $E_F$, consistent with the formal N$^{3-}$ ionic picture. In contrast, the valence region of Ba appears nearly empty, and conduction bands above $E_F$ predominantly originate from unoccupied Ba 6$s$ orbitals. Notably, the sum of the contributions from Ba and N atoms cannot fully account for the total DOS near $E_F$~\cite{zhang2022}. Instead, the contributions from the interstitial sites are significant around $E_F$, which is a defining signature of electrides~\cite{zhang2022, lee2013, zhang2014, zhu2019, mcrae2022}. The PDOS of the interstitial site exhibits a spectral shape similar to that of the Ba PDOS. These overall features again confirm that Ba atoms donate electrons to N atoms and the remaining excess electrons from Ba atoms occupy the interstitial site as discussed above~\cite{rauch1992, snyder1995, stein1998, oliv2005, zhang2022}.

Upon sodiation, intercalated Na atoms systematically donate electrons to the Ba$_3$N framework and interstitial sites. This electron donation induces a rigid-band shift effect, moving $E_F$ progressively upward. As a clear manifestation of this shift, a characteristic band crossing at the K point (highlighted by red circles in Figure~\ref{dos}) is positioned at -0.7 eV relative to $E_F$ in the pristine Ba$_3$N case (Figure~\ref{dos}a). Upon Na intercalation, this band crossing shifts downward to -1 eV for NaBa$_3$N case (Figure~\ref{dos}b) and further to -2 eV for Na$_5$Ba$_3$N case (Figure~\ref{dos}c), respectively. Note that the band structure of orthorhombic Na$_5$Ba$_3$N is unfolded back to the hexagonal Brillouin zone of the pristine Ba$_3$N cell for a direct comparison.

\begin{figure*}
\includegraphics{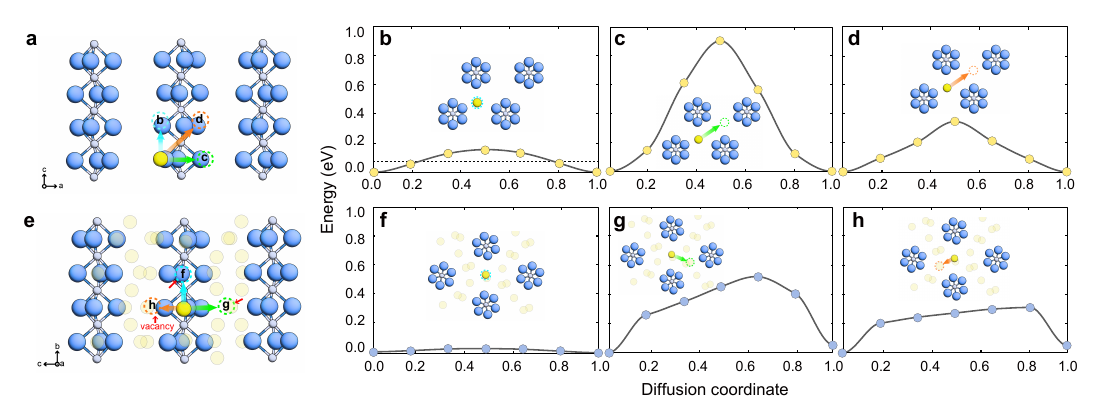}
\caption{Nudged elastic band (NEB) calculations for Na-ion migration pathways in the Ba$_3$N electride framework. (a) Three candidate diffusion pathways for a single Na ion in the Na-dilute limit: intrachain migration along the chain direction (Path 1, cyan), transverse interchain diffusion (Path 2, green), and diagonal cross-chain migration (Path 3, orange). (b-d) Calculated energy profiles along the diffusion coordinate for Path 1 (b), Path 2 (c), and Path 3 (d), with activation barriers of 0.15, 0.90, and 0.34 eV, respectively. (e) Vacancy-assisted Na hopping pathways in the Na-rich limit (Na$_{4.75}$Ba$_3$N): nearest-neighbor hops into a Na vacancy along the chain direction (Path 1, cyan) and across different interchain arrangements (Paths 2 and 3, orange and green). (f-h) Calculated energy profiles for Path 1 (f), Path 2 (g), and Path 3 (h), with activation barriers of 0.03, 0.31, and 0.52 eV, respectively. Blue, gray, and yellow spheres denote Ba, N, and Na atoms; dashed spheres in (e) mark Na vacancy sites.}
\label{NEB}
\end{figure*}

Even upon heavy intercalation of up to five Na atoms, both the electride character and intrinsic metallic nature are fully preserved~\cite{zhang2022, weaver2025}. As observed in the spatial charge density distribution (Figure~\ref{struct}c), the PDOS further confirms that the interstitial site maintains a predominant contribution near $E_F$ across all Na intercalation levels~\cite{zhang2022}. Furthermore, the DOS clearly exhibits a substantial density at $E_F$ throughout the entire sodiation process, demonstrating persistent metallic behavior that satisfies a critical prerequisite for battery anode materials. This intrinsic electronic conductivity enables effective operation without relying on conductive additives.

Figure~\ref{convex}a shows the calculated convex hull of the Na--Ba$_3$N system, evaluating the thermodynamic stability of various intercalated Na$_x$Ba$_3$N phases. NaBa$_3$N and Na$_5$Ba$_3$N lie on the convex hull line, whereas other intermediate phases reside above it. This indicates that Na can spontaneously intercalate into the Ba$_3$N framework to form thermodynamically stable intermediate phases, namely NaBa$_3$N and Na$_5$Ba$_3$N, in excellent agreement with experimental observations and a previous theoretical study~\cite{rauch1992, snyder1995, stein1998, zhang2022}. Notably, the formation energies of these phases are relatively small (around -50 meV/atom), and this shallow energy hull directly governs the low reaction voltage profile advantageous for anode applications.

Figure~\ref{convex}b presents the calculated voltage--capacity profile of Na$_x$Ba$_3$N, overlaid with the experimental voltage curve of a hard carbon anode for SIBs~\cite{weaving2020}. As expected from the shallow convex hull, Na$_x$Ba$_3$N exhibits a remarkably low and flat voltage plateau strictly below 0.2 V relative to Na/Na$^+$. Unlike hard carbon, which shows a pronounced high-voltage sloping region (attributed to initial surface and defect adsorption that degrades the overall energy density of full cells), the Ba$_3$N electride framework maintains a low-voltage reaction throughout the sodiation process. Consequently, Na$_x$Ba$_3$N features a significantly wider low-voltage plateau region (< 0.2 V), delivering a high specific capacity of approximately 250 mAh/g, which offers superior practical energy density for SIB anode materials.

Along with energy density, rate capability is a critical determinant for high-performance battery anodes, where the overall rate performance is governed by both electronic and ionic transport within the host structure. While the intrinsic metallic nature of Na$_x$Ba$_3$N ensures excellent electronic conductivity (Figure~\ref{dos}), the rate capability is fundamentally determined by the ionic mobility of Na$^+$ ions. To evaluate the kinetics of Na transport, we performed nudged elastic band (NEB) calculations to investigate the migration pathways and energy barriers in the Na$_x$Ba$_3$N electride framework under both Na-dilute and Na-rich conditions (Figure~\ref{NEB}).

Figure~\ref{NEB}a illustrates the three candidate diffusion pathways in the Na-dilute limit (NaBa$_3$N), where a single Na ion migrates within the Ba$_3$N electride framework. 
Path 1 (cyan arrow, Figure ~\ref{NEB}b) corresponds to intrachain migration along the 1D Ba$_3$N chain direction over a distance of 3.52 \AA, while Path 2 (green arrow, Figure ~\ref{NEB}c) represents transverse interchain diffusion perpendicular to the chain direction across neighboring chains (4.50 \AA).
Path 3 (orange arrow, Figure ~\ref{NEB}d) denotes diagonal cross-chain migration, where the Na ion hops diagonally across the chains while advancing along the axial direction (5.70 \AA). 
The calculated activation energy barriers for Paths 1, 2, and 3 are 0.15, 0.90, and 0.34 eV, respectively. 
Notably, the barrier for Path 1 (0.15 eV) is less than half of the typical diffusion barrier reported for commercial LIB anodes (e.g., $\sim0.4$ eV for Li$_x$C$_6$)~\cite{persson2010, Takahara2021}, highlighting highly facile 1D Na-ion transport along the chain direction even at the initial sodiation stage.
While the interchain pathway (Path 2) exhibits a considerably higher barrier, the diagonal cross-chain pathway (Path 3) shows a barrier of 0.34 eV, comparable to the Li diffusion barrier in graphite anodes, indicating that interchain Na-ion transport does not preclude viable mobility across the Ba$_3$N framework. 
We additionally examined a more dilute limit at a lower Na content (Na$_{0.5}$Ba$_3$N) and the intrachain migration barrier (Path 1) decreases even further, falling below 0.1 eV (dashed line in Fig. ~\ref{NEB}b).

To further investigate the diffusion kinetics under Na-rich conditions, NEB calculations were performed for the vacancy-assisted hopping mechanism in Na$_{4.75}$Ba$_3$N (Figure \ref{NEB}e). Paths 1, 2, and 3 represent nearest-neighbor hops into a Na vacancy along the chain direction (cyan arrow, Figure~\ref{NEB}f) and across different inter-chain arrangements (orange and green arrows, Figures~\ref{NEB}g and h), with migration distances of 2.62, 2.87, and 4.57 \AA, respectively. The corresponding activation barriers are 0.03, 0.31, and 0.52 eV. Remarkably, the energy barrier along the chain direction drops to an ultra-low value of 0.03 eV, which is virtually barrierless and indicative of superionic-like Na transport. The asymmetry in the energy profiles of Figure~\ref{NEB}g and h reflects the structurally inequivalent local environments at the two endpoints of each hop, arising from the lower local symmetry along these interchain pathways. Overall, these results confirm that Na (de)intercalation in Ba$_3$N proceeds rapidly not only along the 1D chain direction but also across interchain pathways, guaranteeing exceptional high-rate capability for battery applications.

\section{Discussion}

To serve as a high-performance battery anode, a host material fundamentally requires both high electrical and ionic conductivity. Notably, most known inorganic electrides exhibit intrinsic metallic conductivity~\cite{tada2014, zhu2019, mcrae2022, weaver2025}, since the interstitial anionic electrons are delocalized along the open interstitial channels, giving rise to persistent metallic electronic conductivity. Beyond this generic metallicity, Ba$_3$N further exhibits exceptionally rapid Na$^+$ ion diffusion discussed above, and this is not unique to Ba$_3$N alone. A similar behavior has been reported in high-pressure hcp-Fe, where the emergence of an electride phase enables remarkable proton diffusion, believed to underlie the superionic state of iron-rich planetary interiors~\cite{he2022, park2024}. In addition, we find that the prototypical 2D electride Ca$_2$N also exhibits a remarkably low Na$^+$ ion migration barrier of 0.07 eV (see Supporting Information Fig.S2), even though Na intercalation into Ca$_2$N is not thermodynamically favorable~\cite{hu2015}.

\begin{figure*}
	\includegraphics[width=1\textwidth]{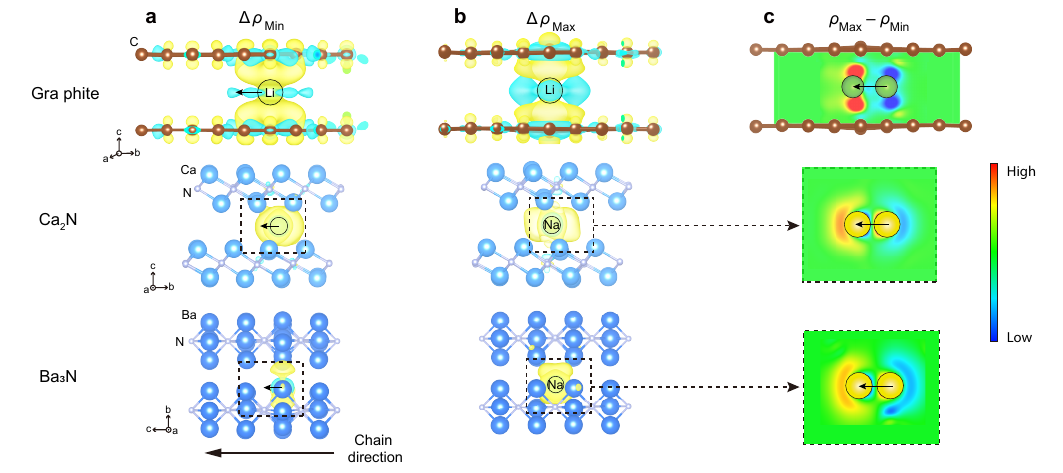}
	\caption{Real-space electronic redistribution and charge transfer characteristics during cation migration. (A, B) Three-dimensional charge density difference plots ($\Delta\rho = \rho_{\text{total}} - \rho_{\text{ion}} - \rho_{\text{host}}$) at the (A) minimum (Min) and (B) transition state (Max) along the NEB migration pathway for Li in graphite, Na in Ca$_2$N, and Na in Ba$_3$N. Yellow and cyan isosurfaces indicate electron accumulation and depletion, respectively. (C) Real-space charge density variation during ion hopping, calculated as $\rho_{\text{Max}} - \rho_{\text{Min}}$. In graphite, the charge perturbation displays pronounced vertical anisotropy extending toward the adjacent carbon rings, reflecting host–ion interactions, whereas both $\mathrm{Ca_2N}$ and $\mathrm{Ba_3N}$ exhibit highly isotropic contours strictly confined around the migrating species.}
	\label{charge}
\end{figure*}

To fundamentally understand this exceptionally low energy barrier for cation migration in electride materials, we analyze the charge density distribution at the energy minimum (initial site) and the energy maximum (transition state) along the NEB migration path as shown in Fig.~\ref{charge} (see Supporting Information Fig.S2 for the graphite-Li and Ca$_2$N-Na NEB case). Specifically, the charge density difference ($\Delta \rho$) at each state was evaluated as:
\begin{equation}
    \label{eq1}
\Delta\rho = \rho_{\text{total}} - \rho_{\text{atom}} - \rho_{\text{host}}
\end{equation}
where $\rho_{\text{total}}$, $\rho_{\text{atom}}$, and $\rho_{\text{host}}$ represent the charge densities of the intercalated systems, the isolated migrating atom (Li for graphite, and Na for Ca$_2$N and Ba$_3$N), and the corresponding bare host framework, respectively (Fig.~\ref{charge}a and b).
Furthermore, to capture the real-space charge redistribution during the hopping process, we mapped the differential charge density between the transition state and the minimum state ($\rho_{\text{Max}} - \rho_{\text{Min}}$), as illustrated in Fig.~\ref{charge}c.

In graphite, the migration of Li directly perturbs the charge distribution of the host framework across all stages of transport. As shown in Fig.~\ref{charge}a and b (top panels), pronounced regions of electron accumulation (yellow) and depletion (cyan) extend notably across the adjacent carbon sheets in both the minimum and maximum states. This indicates substantial charge transfer and orbital hybridization between the diffusing Li species and the host graphite network. Consequently, during the hopping process itself, Li translation induces a substantial anisotropic charge perturbation that extends vertically toward the adjacent carbon sheets ($\rho_{\text{Max}} - \rho_{\text{Min}}$, Fig.~\ref{charge}c, top). This demonstrates that Li hopping is intrinsically coupled to local orbital reorganization and electrostatic interactions with the carbon framework, defining the standard potential energy barrier typical of conventional intercalation hosts.

In stark contrast, the electrides Ca$_2$N and Ba$_3$N display remarkably inert host frameworks throughout the migration process (Fig.~\ref{charge}a and b, middle and bottom panels). The host frameworks, the [Ca$_2$N]$^+$ layers and [Ba$_3$N]$^{3+}$ chains, show negligible changes in their electron density, with virtually no visible accumulation or depletion on the Ca, Ba, or N sites. Instead, charge transfer and redistribution are strictly confined within the open interstitial regions where excess anionic electrons reside. 
This picture is consistent with the previous theoretical study demonstrating that the excess electrons introduced by intercalated Na species in Na$_5$Ba$_3$N mostly populate the interstitial void rather than the host framework~\cite{zhang2022}.
Tracking the real-space variation during hopping ($\rho_{\text{Max}} - \rho_{\text{Min}}$, Fig.~\ref{charge}c) further reveals an exceptionally smooth and isotropic profile centered strictly on the moving Na species, without perturbing the surrounding lattice atoms. Crucially, this profile closely resembles the intrinsic electronic redistribution of an isolated atom translating through free space (see Supporting Information Fig.S3). 

In conventional intercalation hosts like graphite, inserted ions induce substantial electronic reorganization within the host framework, effectively trapping the migrating cations at specific crystallographic sites (e.g., hollow sites). Consequently, ion hopping inherently requires overcoming these localized orbital interactions and escaping the associated potential energy wells, which dictates the standard activation barrier. In electride materials, however, the interstitial space is occupied by a spatially uniform, delocalized anionic electron sea rather than localized bonding orbitals.
Because this compliant electron cloud homogeneously accommodates the migrating cation species along the entire channel, transport proceeds without inducing localized orbital hybridization or site-specific trapping with the host framework.
This absence of rigid local electronic interactions flattens the potential energy landscape, enabling ultrafast transport that underpins the unprecedentedly low migration barrier. This barrierless, fluidic translation aligns with previous findings on layered electrides, which revealed that a homogeneous interstitial electron gas effectively flattens local electrostatic corrugations, thereby inducing ultralow sliding friction with exceptionally small interlayer sliding barriers ($\sim17$ meV)~\cite{wang2018ultralow, druffel2016}.

Finally, the mechanical integrity of Ba$_3$N under substantial sodiation requires practical consideration. Accommodating Na up to the terminal Na$_5$Ba$_3$N phase involves a unit-cell volume expansion of $\sim 100\%$. While this volumetric change is larger than that of conventional carbon hosts, it is substantially more moderate than the severe, catastrophic volume expansions typical of alloying-type anodes (e.g., $>400\%$ for Sn and $>300\%$ for P), which typically suffer from particle pulverization and rapid capacity fading.  

Crucially, the quasi-one-dimensional architecture of Ba$_3$N provides intrinsic structural compliance to accommodate this mechanical strain. The [Ba$_3$N]$^{3+}$ chains are cohesive not through stiff covalent or ionic networks but via delocalized interstitial anionic electrons that act as a compliant electrostatic glue. Although this interchain coupling is electrostatic and metallic in origin rather than a van der Waals (vdW) interaction, it exhibits an extraordinarily soft and flexible character. As demonstrated in the layered electride Ca$_2$N, such electron-mediated interlayer bonding enables exfoliation with a binding energy comparable in magnitude to that of graphite~\cite{zhao2014, druffel2016}, alongside an exceptionally low barrier for interlayer sliding~\cite{wang2018ultralow}, underscoring the inherently soft and shear-compliant nature of inter-unit coupling in electrides.

Furthermore, the interlayer binding potential well for such electrides is remarkably broad along the separation distance compared to conventional vdW materials~\cite{druffel2016}; unlike dipole-induced vdW interactions that decay rapidly with separation, the interaction mediated by the interstitial electron sea remains effective across substantially expanded separations (see Supporting Information Fig.S4 for the comparison between Ba$_3$N and 1D vdW Te chains). More importantly, this structural cohesion is dynamically self-sustaining, as the terminal Na$_5$Ba$_3$N phase itself persists as an electride. With excess electrons from the intercalated Na continuously populating the expanding interstitial channels, an abundant ``electron glue '' is persistently regenerated between the dilated [Ba$_3$N]$^{3+}$ chains. Consequently, the host framework accommodates substantial volumetric expansion ($\sim 100\%$) without interfacial decohesion or lattice collapse, a structural robustness directly confirmed by our room-temperature ab initio molecular dynamics (AIMD) simulations of both the pristine Ba$_3$N and terminal Na$_5$Ba$_3$N phases (see Supporting Information Fig.S5).

\section{Conclusion}
In conclusion, using comprehensive first-principles calculations, we have established the one-dimensional subnitride electride $\mathrm{Ba_3N}$ as a promising, ultra-fast anode material for NIBs. Thermodynamic convex hull evaluations confirm that $\mathrm{Ba_3N}$ spontaneously intercalates sodium up to the terminal $\mathrm{Na_5Ba_3N}$ phase, establishing a flat, low-potential reaction plateau ($<0.2\text{ V}$ vs. $\mathrm{Na/Na}^+$) that delivers a gravimetric capacity of $\sim 250\text{ mAh g}^{-1}$. This reaction profile resolves the trade-off of conventional hard carbon by eliminating high-voltage sloping energy losses while maintaining an essential safety margin against dendrite plating. 
Crucially, the spatially uniform interstitial electron sea acts as a compliant buffer that suppresses local orbital interaction with the host framework, smoothing the diffusion landscape to enable superionic-like Na transport with an ultra-low barrier. Moreover, the non-directional electron glue imparts structural compliance to accommodate the $\sim 100\%$ volume dilation without lattice breakdown. This work establishes a design paradigm for harnessing electride chemistry to achieve ultrafast battery electrodes.

\section{Experimental Section}
All calculations were performed using the Vienna ab initio simulation package (\texttt{VASP}) with the projector-augmented wave (PAW) method~\cite{vasp1,vasp2}. The exchange-correlation interactions were consistently treated with the Perdew–Burke–Ernzerhof (PBE) functional within the generalized gradient approximation (GGA)~\cite{gga}. To account for long-range dispersion forces, various van der Waals (vdW) correction methods were employed, such as the DFT-D2, DFT-D3 (with zero-damping and Becke-Johnson damping), and the nonlocal optB86b-vdW functionals~\cite{vdwd2,d3-zero,d3-becke,vdw-opt}. To determine the most appropriate vdW correction scheme for this system, structural relaxation tests were performed using each method, and the resulting lattice parameters were compared with the experimental lattice parameter of Ba$_3$N (Supporting Information Fig.S1). Among the methods tested, DFT-D3 with zero-damping yielded the lattice parameter in closest agreement with experiment, and was therefore selected for all subsequent calculations in this study. A plane-wave cut-off energy of 500 eV was applied. For electronic structure calculations, The Monkhorst-Pack k-point mesh with a resolution of 0.030/\AA ~was used to determine the density of states (DOS). \\
The diffusion energies were calculated using the nudged elastic band (NEB) method~\cite{henkelman2000}. A spring constant of -5 eV/\AA$^{2}$ ~was set between the images, and the structures were relaxed until the atomic forces were less than 0.05 $eV$.  To avoid spurious interactions between periodic images of the diffusing ions, a minimum separation distance of 7 \AA ~was maintained~\cite{neb_lbfgs,persson2010}. The supercell dimensions were determined based on the formula units (f.u.); an 8-f.u. supercell was employed for both the Na-dilute phase (containing 1 Na atom in a Ba$_{24}$N$_{8}$ framework) and the Na-rich phase (containing 37 Na atoms within the same framework).\\
AIMD simulations were conducted at 300 $K$ to assess the structural integrity and thermal stability of the system over time. These simulations employed a kinetic energy cut-off of 500 eV and the Nosé-Hoover thermostat to maintain the canonical ensemble~\cite{Douglas1993}. The total simulation duration was 8 $ps$.\\

\begin{acknowledgements}
The authors thank Min-Sik Park, Beomjoon Goh, and Ina Park for fruitful discussions.
This work is supported by the National Research Council of Science \& Technology (NST) grant by the Korea government (MSIT) (No. CAP25061-000), 
a National Research Foundation (NRF) of Korea grantfunded by the Korea government (MSIT) (RS-2026-25488595),
The computational resources were provided by the Center for Advanced Computation (CAC) at Korea Institute for Advanced Study (KIAS) and the National Supercomputing Center, along with technical support (KSC-2025-CRE-0367).
\end{acknowledgements}

\bibliography{Electride2}

@article{stevens2000high,
  title={High capacity anode materials for rechargeable sodium-ion batteries},
  author={Stevens, DA and Dahn, JR},
  journal={Journal of the Electrochemical Society},
  volume={147},
  number={4},
  pages={1271--1273},
  year={2000},
  publisher={The Electrochemical Society, Inc.}
}

@article{yabuuchi2014,
  title={Research development on sodium-ion batteries},
  author={Yabuuchi, Naoaki and Kubota, Kei and Dahbi, Mouad and Komaba, Shinichi},
  journal={Chemical Reviews},
  volume={114},
  number={23},
  pages={11636--11682},
  year={2014},
  publisher={ACS Publications}
}

@article{komaba2011,
  title={Electrochemical Na insertion and solid electrolyte interphase for hard-carbon electrodes and application to Na-Ion batteries},
  author={Komaba, Shinichi and Murata, Wataru and Ishikawa, Toru and Yabuuchi, Naoaki and Ozeki, Tomoaki and Nakayama, Tetsuri and Ogata, Atsushi and Gotoh, Kazuma and Fujiwara, Kazuya},
  journal={Advanced Functional Materials},
  volume={21},
  number={20},
  pages={3859--3867},
  year={2011},
  publisher={Wiley Online Library}
}

@article{xiao2018low,
  title={Low-defect and low-porosity hard carbon with high coulombic efficiency and high capacity for practical sodium ion battery anode},
  author={Xiao, Lifen and Lu, Haiyan and Fang, Yongjin and Sushko, Maria L and Cao, Yuliang and Ai, Xinping and Yang, Hanxi and Liu, Jun},
  journal={Advanced Energy Materials},
  volume={8},
  number={20},
  pages={1703238},
  year={2018},
  publisher={Wiley Online Library}
}

@article{qi2019slope,
  title={Slope-dominated carbon anode with high specific capacity and superior rate capability for high safety Na-ion batteries},
  author={Qi, Yuruo and Lu, Yaxiang and Ding, Feixiang and Zhang, Qiangqiang and Li, Hong and Huang, Xuejie and Chen, Liquan and Hu, Yong-Sheng},
  journal={Angewandte Chemie International Edition},
  volume={58},
  number={13},
  pages={4361--4365},
  year={2019},
  publisher={Wiley Online Library}
}

@article{wang2024recent,
  title={Recent progress in hard carbon anodes for sodium-ion batteries},
  author={Wang, Jiarui and Xi, Lei and Peng, Chenxi and Song, Xin and Wan, Xuanhong and Sun, Luyi and Liu, Meinan and Liu, Jun},
  journal={Advanced Engineering Materials},
  volume={26},
  number={8},
  pages={2302063},
  year={2024},
  publisher={Wiley Online Library}
}

@article{bommier2015new,
  title={New mechanistic insights on Na-ion storage in nongraphitizable carbon},
  author={Bommier, Clement and Surta, Todd Wesley and Dolgos, Michelle and Ji, Xiulei},
  journal={Nano Letters},
  volume={15},
  number={9},
  pages={5888--5892},
  year={2015},
  publisher={ACS Publications}
}

@article{zhang2020extended,
  title={Extended low-voltage plateau capacity of hard carbon spheres anode for sodium ion batteries},
  author={Zhang, Xiang and Dong, Xiaoli and Qiu, Xuan and Cao, Yongjie and Wang, Congxiao and Wang, Yonggang and Xia, Yongyao},
  journal={Journal of Power Sources},
  volume={476},
  pages={228550},
  year={2020},
  publisher={Elsevier}
}

@article{li2017mechanism,
  title={Mechanism of Na-ion storage in hard carbon anodes revealed by heteroatom doping},
  author={Li, Zhifei and Bommier, Clement and Chong, Zhi Sen and Jian, Zelang and Surta, Todd Wesley and Wang, Xingfeng and Xing, Zhenyu and Neuefeind, Joerg C and Stickle, William F and Dolgos, Michelle and others},
  journal={Advanced Energy Materials},
  volume={7},
  number={18},
  pages={1602894},
  year={2017},
  publisher={Wiley Online Library}
}

@article{zhao2014,
  title={Obtaining two-dimensional electron gas in free space without resorting to electron doping: an electride based design},
  author={Zhao, Songtao and Li, Zhenyu and Yang, Jinlong},
  journal={Journal of the American Chemical Society},
  volume={136},
  number={38},
  pages={13313--13318},
  year={2014},
  publisher={ACS Publications}
}

@article{druffel2016,
  title={Experimental demonstration of an electride as a 2D material},
  author={Druffel, Daniel L and Kuntz, Kaci L and Woomer, Adam H and Alcorn, Francis M and Hu, Jun and Donley, Carrie L and Warren, Scott C},
  journal={Journal of the American Chemical Society},
  volume={138},
  number={49},
  pages={16089--16094},
  year={2016},
  publisher={ACS Publications}
}

@article{wang2018ultralow,
  title={Ultralow interlayer friction of layered electride Ca2N: A potential two-dimensional solid lubricant material},
  author={Wang, Jianjun and Li, Lin and Shen, Ziting and Guo, Peng and Li, Meng and Zhao, Bin and Fang, Lili and Yang, Linfeng},
  journal={Materials},
  volume={11},
  number={12},
  pages={2462},
  year={2018},
  publisher={MDPI}
}

@article{dawes1986first,
  title={First electride crystal structure},
  author={Dawes, Steven B and Ward, Donald L and Huang, Rui He and Dye, James L},
  journal={Journal of the American Chemical Society},
  volume={108},
  number={12},
  pages={3534--3535},
  year={1986},
  publisher={ACS Publications}
}

@article{li2003inorganic,
  title={Inorganic electride: theoretical study on structural and electronic properties},
  author={Li, Zhenyu and Yang, Jinlong and Hou, JG and Zhu, Qingshi},
  journal={Journal of the American Chemical Society},
  volume={125},
  number={20},
  pages={6050--6051},
  year={2003},
  publisher={ACS Publications}
}

@article{redko2005design,
  title={Design and synthesis of a thermally stable organic electride},
  author={Redko, Mikhail Y and Jackson, James E and Huang, Rui H and Dye, James L},
  journal={Journal of the American Chemical Society},
  volume={127},
  number={35},
  pages={12416--12422},
  year={2005},
  publisher={ACS Publications}
}

@article{liu2020electrides,
  title={Electrides: a review},
  author={Liu, Chang and Nikolaev, Sergey A and Ren, Wei and Burton, Lee A},
  journal={Journal of Materials Chemistry C},
  volume={8},
  number={31},
  pages={10551--10567},
  year={2020},
  publisher={The Royal Society of Chemistry}
}

@article{hosono2021advances,
  title={Advances in materials and applications of inorganic electrides},
  author={Hosono, Hideo and Kitano, Masaaki},
  journal={Chemical Reviews},
  volume={121},
  number={5},
  pages={3121--3185},
  year={2021},
  publisher={ACS Publications}
}

@article{zhou2024van,
  title={Van der Waals electrides},
  author={Zhou, Jun and You, Jing-Yang and Zhao, Yi-Ming and Feng, Yuan Ping and Shen, Lei},
  journal={Accounts of Chemical Research},
  volume={57},
  number={17},
  pages={2572--2581},
  year={2024},
  publisher={ACS Publications}
}

@article{hu2015,
  title={{2D electrides as promising anode materials for Na-ion batteries from first-principles study}},
  author={Hu, Junping and Xu, Bo and Yang, Shengyuan A and Guan, Shan and Ouyang, Chuying and Yao, Yugui},
  journal={ACS Applied Materials \& Interfaces},
  volume={7},
  number={43},
  pages={24016--24022},
  year={2015},
  publisher={ACS Publications}
}

@article{he2022,
  title={{Superionic iron alloys and their seismic velocities in Earth's inner core}},
  author={He, Yu and Sun, Shichuan and Kim, Duck Young and Jang, Bo Gyu and Li, Heping and Mao, Ho-kwang},
  journal={Nature},
  volume={602},
  number={7896},
  pages={258--262},
  year={2022},
  publisher={Nature Publishing Group UK London}
}

@article{tada2014,
  title={{High-throughput ab initio screening for two-dimensional electride materials}},
  author={Tada, Tomofumi and Takemoto, Seiji and Matsuishi, Satoru and Hosono, Hideo},
  journal={Inorganic Chemistry},
  volume={53},
  number={19},
  pages={10347--10358},
  year={2014},
  publisher={ACS Publications}
}

@article{zhang2014,
  title={{Two-dimensional transition-metal electride Y$_{2}$C}},
  author={Zhang, Xiao and Xiao, Zewen and Lei, Hechang and Toda, Yoshitake and Matsuishi, Satoru and Kamiya, Toshio and Ueda, Shigenori and Hosono, Hideo},
  journal={Chemistry of Materials},
  volume={26},
  number={22},
  pages={6638--6643},
  year={2014},
  publisher={ACS Publications}
}

@article{mcrae2022,
  title={{Sc$_{2}$C, a 2D semiconducting electride}},
  author={McRae, Lauren M and Radomsky, Rebecca C and Pawlik, Jacob T and Druffel, Daniel L and Sundberg, Jack D and Lanetti, Matthew G and Donley, Carrie L and White, Kelly L and Warren, Scott C},
  journal={Journal of the American Chemical Society},
  volume={144},
  number={24},
  pages={10862--10869},
  year={2022},
  publisher={ACS Publications}
}

@article{weaver2025,
  title={{Assessing dimensionality in electrides}},
  author={Weaver, Samuel M and Lanetti, Matthew G and Slamowitz, Connor C and Radomsky, Rebecca C and Warren, Scott C},
  journal={The Journal of Physical Chemistry C},
  volume={129},
  number={6},
  pages={3211--3217},
  year={2025},
  publisher={ACS Publications}
}

@article{komaba2009,
  title={{Electrochemically reversible sodium intercalation of layered NaNi$_{0.5}$Mn$_{0.5}$O$_{2}$ and NaCrO$_{2}$}},
  author={Komaba, Shinichi and Nakayama, Tetsuri and Ogata, Atsushi and Shimizu, Takaya and Takei, Chikara and Takada, S and Hokura, A and Nakai, I},
  journal={ECS Transactions},
  volume={16},
  number={42},
  pages={43},
  year={2009},
  publisher={IOP Publishing}
}

@article{komaba2010,
  title={{Electrochemical intercalation activity of layered NaCrO$_{2}$ vs. LiCrO$_{2}$}},
  author={Komaba, Shinichi and Takei, Chikara and Nakayama, Tetsuri and Ogata, Atsushi and Yabuuchi, Naoaki},
  journal={Electrochemistry Communications},
  volume={12},
  number={3},
  pages={355--358},
  year={2010},
  publisher={Elsevier}
}

@article{liu2023,
  title={{Na-Rich Na$_{3}$V$_{2}$(PO$_{4}$)$_{3}$ Cathodes for Long Cycling Rechargeable Sodium Full Cells}},
  author={Liu, Yao and Wu, Xiangyong and Moeez, Abdul and Peng, Zhi and Xia, Yongyao and Zhao, Dongyuan and Liu, Jun and Li, Wei},
  journal={Advanced Energy Materials},
  volume={13},
  number={3},
  pages={2203283},
  year={2023},
  publisher={Wiley Online Library}
}

@article{hurlbutt2018,
  title={{Prussian blue analogs as battery materials}},
  author={Hurlbutt, Kevin and Wheeler, Samuel and Capone, Isaac and Pasta, Mauro},
  journal={Joule},
  volume={2},
  number={10},
  pages={1950--1960},
  year={2018},
  publisher={Elsevier}
}

@article{moriwake2017,
  title={{Why is sodium-intercalated graphite unstable?}},
  author={Moriwake, Hiroki and Kuwabara, Akihide and Fisher, Craig AJ and Ikuhara, Yuichi},
  journal={RSC Advances},
  volume={7},
  number={58},
  pages={36550--36554},
  year={2017},
  publisher={Royal Society of Chemistry}
}

@article{dye1990,
  title={{Electrides: ionic salts with electrons as the anions}},
  author={Dye, James L},
  journal={Science},
  volume={247},
  number={4943},
  pages={663--668},
  year={1990},
  publisher={American Association for the Advancement of Science}
}

@article{dye2009,
  title={{Electrides: early examples of quantum confinement}},
  author={Dye, James L},
  journal={Accounts of Chemical Research},
  volume={42},
  number={10},
  pages={1564--1572},
  year={2009},
  publisher={ACS Publications}
}

@article{zhu2019,
  title={{Computational discovery of inorganic electrides from an automated screening}},
  author={Zhu, Qiang and Frolov, Timofey and Choudhary, Kamal},
  journal={Matter},
  volume={1},
  number={5},
  pages={1293--1303},
  year={2019},
  publisher={Elsevier}
}

@article{lee2013,
  title={{Dicalcium nitride as a two-dimensional electride with an anionic electron layer}},
  author={Lee, Kimoon and Kim, Sung Wng and Toda, Yoshitake and Matsuishi, Satoru and Hosono, Hideo},
  journal={Nature},
  volume={494},
  number={7437},
  pages={336--340},
  year={2013},
  publisher={Nature Publishing Group UK London}
}

@article{park2024,
  title={{Electride Formation of HCP-Iron at High Pressure: Unraveling the Origin of the Superionic State of Iron-Rich Compounds in Rocky Planets}},
  author={Park, Ina and He, Yu and Mao, Ho-kwang and Shim, Ji Hoon and Kim, Duck Young},
  journal={Advanced Science},
  volume={11},
  number={24},
  pages={2308177},
  year={2024},
  publisher={Wiley Online Library}
}

@article{rauch1992,
  author={Rauch, Paul E. and Simon, Arndt},
  title={{The New Subnitride NaBa$_{3}$N; an Extension of Alkali Metal Suboxide Chemistry}},
  journal={Angewandte Chemie International Edition},
  volume={31},
  number={11},
  pages={1519--1521},
  year={1992}
}

@article{snyder1995,
  author={Snyder, G. Jeffrey and Simon, Arndt},
  title={{The Infinite Chain Nitride Na$_{5}$Ba$_{3}$N. A One-Dimensional Void Metal?}},
  journal={Journal of the American Chemical Society},
  volume={117},
  number={7},
  pages={1996--1999},
  year={1995}
}

@article{stein1998,
  author={Steinbrenner, U. and Simon, A.},
  title={{Ba$_{3}$N – a New Binary Nitride of an Alkaline Earth Metal}},
  journal={Zeitschrift f{\"u}r anorganische und allgemeine Chemie},
  volume={624},
  number={2},
  pages={228--232},
  year={1998}
}

@article{oliv2005,
  title={{Subnitride chemistry: A first-principles study of the NaBa$_{3}$N, Na$_{5}$Ba$_{3}$N, and Na$_{16}$Ba$_{6}$N phases}},
  journal={Journal of Solid State Chemistry},
  volume={178},
  number={4},
  pages={1023--1029},
  year={2005},
  author={Oliva, Josep M.}
}

@article{zhang2022,
  author={Zhang, Xiangyu and Chen, Yunlei and Sun, Yongfang and Ye, Tian-Nan and Wen, Xiao-Dong},
  title={{First-Principles Study of Three-Dimensional Electrides Containing One-Dimensional [Ba$_{3}$N]$^{3+}$ Chains}},
  journal={ACS Omega},
  volume={7},
  number={15},
  pages={13290--13298},
  year={2022}
}

@article{Douglas1993,
  author={Tobias, Douglas J. and Martyna, Glenn J. and Klein, Michael L.},
  title={{Molecular dynamics simulations of a protein in the canonical ensemble}},
  journal={The Journal of Physical Chemistry},
  volume={97},
  number={49},
  pages={12959--12966},
  year={1993}
}

@article{henkelman2000,
  author={Henkelman, Graeme and J{\'o}nsson, Hannes},
  title={{Improved tangent estimate in the nudged elastic band method for finding minimum energy paths and saddle points}},
  journal={The Journal of Chemical Physics},
  volume={113},
  number={22},
  pages={9978--9985},
  year={2000}
}

@article{vasp1,
  title={{Efficient iterative schemes for ab initio total-energy calculations using a plane-wave basis set}},
  author={Kresse, G. and Furthm{\"u}ller, J.},
  journal={Physical Review B},
  volume={54},
  number={16},
  pages={11169--11186},
  year={1996}
}

@article{vasp2,
  title={{From ultrasoft pseudopotentials to the projector augmented-wave method}},
  author={Kresse, G. and Joubert, D.},
  journal={Physical Review B},
  volume={59},
  number={3},
  pages={1758--1775},
  year={1999}
}

@article{gga,
  title={{Generalized Gradient Approximation Made Simple}},
  author={Perdew, John P. and Burke, Kieron and Ernzerhof, Matthias},
  journal={Physical Review Letters},
  volume={78},
  number={7},
  pages={1396--1396},
  year={1997}
}

@article{vdwd2,
  author={Grimme, Stefan},
  title={{Semiempirical GGA-type density functional constructed with a long-range dispersion correction}},
  journal={Journal of Computational Chemistry},
  volume={27},
  number={15},
  pages={1787--1799},
  year={2006}
}

@article{d3-becke,
  author={Grimme, Stefan and Ehrlich, Stephan and Goerigk, Lars},
  title={{Effect of the damping function in dispersion corrected density functional theory}},
  journal={Journal of Computational Chemistry},
  volume={32},
  number={7},
  pages={1456--1465},
  year={2011}
}

@article{d3-zero,
  author={Grimme, Stefan and Antony, Jens and Ehrlich, Stephan and Krieg, Helge},
  title={{A consistent and accurate ab initio parametrization of density functional dispersion correction (DFT-D) for the 94 elements H-Pu}},
  journal={The Journal of Chemical Physics},
  volume={132},
  number={15},
  pages={154104},
  year={2010}
}

@article{vdw-opt,
  title={{Van der Waals density functionals applied to solids}},
  author={Klime{\v{s}}, Ji{\v{r}}{\'\i} and Bowler, David R. and Michaelides, Angelos},
  journal={Physical Review B},
  volume={83},
  number={19},
  pages={195131},
  year={2011}
}

@article{neb_lbfgs,
  author={Sheppard, Daniel and Terrell, Rye and Henkelman, Graeme},
  title={{Optimization methods for finding minimum energy paths}},
  journal={The Journal of Chemical Physics},
  volume={128},
  number={13},
  pages={134106},
  year={2008}
}

@article{persson2010,
  title={{Thermodynamic and kinetic properties of the Li-graphite system from first-principles calculations}},
  author={Persson, Kristin and Hinuma, Yoyo and Meng, Ying Shirley and Van der Ven, Anton and Ceder, Gerbrand},
  journal={Physical Review B},
  volume={82},
  number={12},
  pages={125416},
  year={2010}
}

@article{Takahara2021,
  title={First principles study on formation and migration energies of sodium and lithium in graphite},
  author={Takahara, Izumi and Mizoguchi, Teruyasu},
  journal={Physical Review Materials},
  volume={5},
  number={8},
  pages={085401},
  year={2021}
}

@article{weaving2020,
  author={Weaving, Julia S. and Lim, Alvin and Millichamp, Jason and Neville, Tobias P. and Ledwoch, Daniela and Kendrick, Emma and McMillan, Paul F. and Shearing, Paul R. and Howard, Christopher A. and Brett, Dan J. L.},
  title={{Elucidating the Sodiation Mechanism in Hard Carbon by Operando Raman Spectroscopy}},
  journal={ACS Applied Energy Materials},
  volume={3},
  number={8},
  pages={7474--7484},
  year={2020}
}

@article{li2015,
  title={{Tin and tin compounds for sodium ion battery anodes: phase transformations and performance}},
  author={Li, Zhi and Ding, Jia and Mitlin, David},
  journal={Accounts of Chemical Research},
  volume={48},
  number={6},
  pages={1657--1665},
  year={2015}
}

@article{jung2016,
  title={{Advantages of Ge anode for Na-ion batteries: Ge vs. Si and Sn}},
  author={Jung, Sung Chul and Kim, Hyung-Jin and Kang, Yong-Ju and Han, Young-Kyu},
  journal={Journal of Alloys and Compounds},
  volume={688},
  pages={158--163},
  year={2016}
}

@article{jung2014,
  title={{Atom-level understanding of the sodiation process in silicon anode material}},
  author={Jung, Sung Chul and Jung, Dae Soo and Choi, Jang Wook and Han, Young-Kyu},
  journal={The Journal of Physical Chemistry Letters},
  volume={5},
  number={7},
  pages={1283--1288},
  year={2014}
}

@article{ni2018,
  title={{Phosphorus: an anode of choice for sodium-ion batteries}},
  author={Ni, Jiangfeng and Li, Liang and Lu, Jun},
  journal={ACS Energy Letters},
  volume={3},
  number={5},
  pages={1137--1144},
  year={2018}
}

\end{document}